\documentclass[a4paper]{article}
\usepackage[english]{babel}
\usepackage[utf8x]{inputenc}
\usepackage[T1]{fontenc}
\usepackage[a4paper,top=3cm,bottom=2cm,left=2.7cm,right=2.7cm,marginparwidth=1.75cm]{geometry}
\usepackage{comment}
\usepackage{authblk}
\usepackage[misc]{ifsym}
\usepackage{indentfirst}
\usepackage[english]{babel}
\usepackage[utf8x]{inputenc}
\usepackage{amsmath,amsfonts,amssymb,amsthm}
\usepackage[english]{babel}
\usepackage{graphicx}
\usepackage{hyperref}
\usepackage{latexsym}
\usepackage{ifthen}
\usepackage{mathtools}
\usepackage{braket}
\usepackage{algorithm}
\usepackage{algpseudocode}
\usepackage{verbatim}
\usepackage[qm]{qcircuit}
\usepackage{todonotes}
\usepackage{multirow}
\usepackage[toc,page]{appendix}
\usepackage{listings}
\usepackage{xcolor}
\usepackage[nottoc,numbib]{tocbibind}
\usepackage{cprotect}
\usepackage{etex}

\makeatletter

\lstdefinelanguage{AMPL}{keywords={set,param,var,arc,integer,minimize,maximize,subject,to,node,sum,in,Current,complements,integer,solve_result_num,IN,contains,less,suffix,INOUT,default,logical,sum,Infinity,dimen,max,symbolic
,Initial,div,min,table,LOCAL,else,option,then,OUT,environ,setof ,union,all,exists,shell_exitcodeuntil,binary,forall,solve_exitcodewhile ,by,if,solve_messagewithin,check,in,solve_result
},sensitive=true,comment=[l]{\#}}

\theoremstyle{plain}
\newtheorem{thm}{Theorem}[section]
\newtheorem{lem}[thm]{Lemma}
\newtheorem{prop}[thm]{Proposition}
\newtheorem{cor}[thm]{Corollary}
\newtheorem{defi}[thm]{Definition}

\title{Quantum algorithms for the Sylvester denumerant and the numerical semigroup membership problem}
\author[1]{J. Ossorio-Castillo\thanks{Email: jqnssr@gmail.com (\Letter). Supported by MTM2016-75027-P (MEyC).}}
\author[1]{Jos\'e M. Tornero \thanks{Email: tornero@us.es. Supported by MTM2016-75027-P (MEyC) and P12-FQM-2696 (JdA).}}
\affil[1]{\textit{IMUS \& Departamento de \'Algebra, Universidad de Sevilla, 41012 Sevilla, Spain}}
\date{}

\begin{document}
\maketitle

\begin{abstract}
Two quantum algorithms are presented, which tackle well--known problems in the context of numerical semigroups: the numerical semigroup membership problem (NSMP) and the Sylvester denumerant problem (SDP). 
\end{abstract}

\section{Numerical Semigroups}\label{sec:semigroups}

Numerical semigroups are essentially additive submonoids of ${\mathbb Z}_{\geq 0}$ with finite complement (see below for a less sophisticated equivalent definition). Being relatively straightforward structures, they accomodate a vast amount of challenging problems whose classical complexity is well--established. \\

Our main target in this paper is to study two computationally hard problems, connected with the theory of numerical semigroups, from the perspective of quantum computation: the Sylvester denumerant problem (SDP) and the numerical semigroup membership problem (NSMP). For both of them we present an algorithm which returns the corresponding solution.\\

These algorithms lean on the quantum circuit model and are based respectively on Grover's database search \cite{grover1996fast} and on quantum counting \cite{brassard1998quantum}. Both of them rely on a generation of all elements of the numerical semigroup up to a certain bound inside the quantum computer thanks to quantum parallelism. This way, we can ask a certain type of questions via an oracle to the distribution of the elements of the numerical semigroup, and then infer the solution to these combinatorial problems. \\

Along with a description of the algorithms, we present some tables of numerical results and hypothetical performance over the classical version, simulated with a \verb|C++| library developed alongside this research. This library is called \verb|numsem| and can be found at a public GitHub repository \cite{ossoriocastillo2018numsem} along with all the documentation needed for the correct replication of the results here presented. \\

It is worth mentioning that this study is a follow-up to a previous work on the subject \cite{ossorio2019adiabatic} made from the perspective of an alternative paradigm of quantum computation: the adiabatic model. In that work, an adiabatic quantum algorithm was presented for the Frobenius problem, well connected to the ones we present here and which we proceed to enunciate.

\begin{defi}\label{def:fcp}
Let $a_1, a_2,\ldots , a_n \in \mathbb{Z}_{\geq 0}$ with $\gcd (a_1, a_2,\ldots , a_n) = 1$, the Frobenius problem, or {\em FP}, is the problem of finding the largest positive integer that cannot be expressed as an integer conical combination of these numbers, i.e., as a sum
$$
\sum_{i=1}^{n} \lambda_i a_i \text{ with } \lambda_i \in \mathbb{Z}_{\geq 0}.
$$
\end{defi}

Focus on this problem from the current perspective started in the 19th Century, mostly by mathematicians J. J. Sylvester and F. G. Frobenius, and gave us the following definition along with a useful characterization.

\begin{defi}
A numerical semigroup $S$ is a subset of the non-negative integers $\mathbb{Z}_{\geq 0}$ which is closed under addition, contains the identity element $0$, and has a finite complement in $\mathbb{Z}_{\geq 0}$.
\end{defi}

\begin{lem}
Let $A = \{a_1,...,a_n\}$ be a nonempty subset of $\mathbb{Z}_{\geq 0}$. Then, 
$$
S = \langle A \rangle = \langle a_1,...,a_n \rangle = \{ \lambda_1a_1 + ... + \lambda_n a_n \ | \ \lambda_i \in \mathbb{Z}_{\geq 0}  \}
$$ 
is a numerical semigroup if and only if $\gcd(a_1, ... , a_n) = 1$.
\end{lem}

From now on, we shall denote numerical semigroups as $S$, taking for granted that they are commutative and that their associated operation is the addition. The concepts and proofs for all the results in this introductory section can be found in \cite{rosales2009numerical} and  \cite{ramirezalfonsin2005diophantine}. \\

By the previous lemma, we can generate a semigroup $S = \langle A \rangle$  from any set $A \subseteq \mathbb{Z}_{\geq 0}$ whose elements satisfy the aforementioned condition. The set $A$ such that $S = \langle A \rangle$ will be called a system of generators of $S$, and it can be proved that any numerical semigroup can be expressed that way.

\begin{thm}
Every numerical semigroup $S$ admits a unique minimal system of generators, which can be calculated as $S^* \setminus (S^* + S^*)$ with $S^* = S \setminus \{0\}$.
\end{thm}

\begin{cor}
Let $S$ be a numerical semigroup generated by $A = \{ a_1 , \ldots , a_n \}$ with $ 0 \neq a_1 < a_2 < ... < a_n $. Then, $A$ is a minimal system of generators of $S$ if and only if $a_{i+1} \notin \langle a_1,a_2,\ldots,a_i \rangle$, for all $i \in \{1, \ldots , n-1 \}$.
\end{cor}

\begin{prop}
The minimal system of generators of a numerical semigroup $S$ is finite.
\end{prop}

Hereinafter, if we say that $S = \langle A \rangle$ with $A = \{ a_1 , a_2 , \ldots , a_n \}$ is a numerical semigroup, then we shall assume without loss of generality that $a_1 < a_2 < \cdots < a_n$, $\gcd (a_1, a_2, \ldots, a_n) = 1$, and that $A$ is the minimal system of generators of $S$. \\

Now we exemplify the relationship between numerical semigroups and combinatorial optimization, as one of the most important problems in the latter branch of mathematics, known as the knapsack problem or rucksack problem, and more concretely one of its variants \cite{papadimitriou1998combinatorial} (p. 374), can be seen as the problem of deciding if a given integer $t$ belongs to a certain numerical semigroup $S$. 

\begin{defi}\label{def:ikp}
The numerical semigroup membership problem, or {\em NSMP}, is the problem of determining if, given a certain integer $t \in \mathbb{Z}_{\geq 0}$ and a numerical semigroup $S = \langle a_1,...,a_n \rangle$, the integer $t$ is contained in $S$. That is to say, if there exist non-negative integers $\lambda_1 , \ldots , \lambda_n \in \mathbb{Z}_{\geq 0}$ such that 
$$
\sum_{i=1}^{n} \lambda_i a_i = t.
$$
\end{defi}

Finally, we show another important problem in numerical semigroups related to the previous one. In the mid-19th century, J. J. Sylvester studied the number of partitions of an integer with respect to a certain subset of non-negative integers \cite{sylvester1857partition,sylvester1896outlines}. In the context of this paper, this problem can be seen as an extension of the NSMP, but rather than answering whether or not an integer is contained in a numerical semigroup, we go farther and want to calculate the number of distinct representations of that integer with respect to the minimal system of generators of the semigroup. This problem will subsequently be called the Sylvester denumerant problem (SDP).

\begin{defi}\label{def:sylden}
The Sylvester denumerant of a non-negative integer $t \in \mathbb{Z}_{\geq 0}$ with respect to a numerical semigroup $S = \langle a_1,...,a_n \rangle$, denoted by $d(t,S)$ or by $d(t;a_1,\ldots,a_n)$, is defined as the number of solutions of the Diophantine equation 
$$
\sum_{i=1}^{n} \lambda_i a_i = t,
$$ 
where $\lambda_1 , \ldots , \lambda_n \in \mathbb{Z}_{\geq 0}$.
\end{defi}

By means of an example, we define the following semigroup:
$$
S = \langle 5,7,9 \rangle = \{0,5,7,9,10,12,14,\rightarrow \}
$$

The Frobenius number of $S$ (the maximum of $\mathbb{Z}_{\geq 0} \setminus S$) is equal to $f(S) = 13$, which means that any integer greater than $13$ is contained in the semigroup (we have noted this in the example with the use of $\rightarrow$, as it is customary in numerical semigroup theory). However, the number of possible representations may differ between them. Number 15, for example, has a unique representation with respect to 5, 7 and 9: 
$$
15 = 3 \times (5) + 0 \times (7)+0 \times (9),
$$ 
while on the other hand, number 14 has two possible representations:
$$
14 = 1 \times (5)+0 \times (7)+1 \times (9)
$$ 
and 
$$
14 = 0 \times (5)+2 \times (7)+0 \times (9).
$$ 
This tells us that $d(15; 5,7,9) = 1 $ and $d(14; 5,7,9) = 2$. \\

The study of the Sylvester denumerant is of great importance in many branches of mathematics and, as can be easily deduced, is at least as hard to solve as the numerical semigroup membership problem. In fact:

\begin{thm}\label{nsmp-np-complete}
The NSMP is in NP--complete.
\end{thm}

\begin{cor}
The SDP is in NP--hard.
\end{cor}

The proof of Theorem \ref{nsmp-np-complete} is well known (or, rather, the one where it is formulated as the knapsack problem) and can be found, for example, in \cite{papadimitriou1998combinatorial}.

\cprotect\section{The \verb|numsem| Library}

In this section we give a brief description of the \verb|numsem| library \cite{ossoriocastillo2018numsem}, specifically implemented for numerical semigroups and which helps us to classically compute the combinatorial invariants described in this paper and also replicate the numerical analyses presented in this chapter. The \verb|numsem| library has been programmed in \verb|C++| and has been tested in Ubuntu 14.04.2. In order to install it, we have just to clone the repository by using the \verb|git| command (provided that we have previously installed \verb|git| inside our computer): \\

\begin{lstlisting}[language=AMPL]
 git clone https://github.com/jqnoc/numsem.git
\end{lstlisting}

After we have downloaded the contents from the repository, we have to type the following installation commands:\\

\begin{lstlisting}[language=AMPL]
 make
 cd bin
 numsem
\end{lstlisting}

Thus, we end up in the directory where the actual executable file is. Let us suppose that we want to test the numerical semigroup $S = \langle 7,11,17,23 \rangle$ or obtain its combinatorial invariants.  We have first to create a file with the generators (we can also just put random integers inside, and let \verb|numsem| decide whether they generate a numerical semigroup or not). In this case, the contents of the file should be like this (the order does not matter):

\begin{lstlisting}[language=AMPL]
 7
 11
 17
 23
\end{lstlisting}

Then, if the file with the generators is called \verb|generators.txt|, we just have to write the following command: 

\begin{lstlisting}[language=AMPL]
 numsem generators.txt
\end{lstlisting}

and obtain the output:

\begin{lstlisting}[language=AMPL]
 ===================================================================
  NumSem v0.1.0
  Copyright (C) Joaquin Ossorio-Castillo. All rights reserved.
  This software is free for non-commercial purposes.
  Full license information can be found in the LICENSE.txt file.
  https://github.com/jqnoc/numsem/
 ===================================================================
 Numerical semigroup: S = <7, 11, 17, 23>
 Total number of generators: 4
\end{lstlisting}

The correct use of the library is triggered with the option \verb|--help|, and also when the options are not written correctly. To date, the available options are:

\begin{lstlisting}[language=AMPL]
 Use:
    numsem *file* [options] 
    *file*: the file with the generators of the numerical semigroup S

 Available options:
    --help: print the help
    -f: calculate Frobenius number of S
    -ga: calculate set of gaps of S
    -gn: calculate genus of S
    -ap *s*: calculate the Apery set of s with respect to S (only with AMPL)
    -d *t*: calculate Sylvester denumerant for t and S
    -ds *t*: calculate Sylvester denumerant for t and S and print all solutions
    -dg *b*: calculate Sylvester function from 0 to b
    -m *t*: calculate if t is in S
    -ampl: use AMPL and Gurobi for Frobenius, genus or Apery set
    -json: write results to a .json file
\end{lstlisting}

We now proceed to explain some of them. The set of gaps is actually $\mathbb{Z}_{\geq 0} \setminus S$. The calculation of the Frobenius number, introduced above, has been implemented in two ways. The default one is quite straightforward and hardly efficient, as it just simply finds the maximum of the set of gaps. The second one is only applied if the option \verb|-ampl| is also activated, and will use one of the algorithms proposed in \cite{ossorio2019adiabatic}. This second option is more efficient and same goes for the genus of $S$ and the Ap\'ery set (which we have not introduced as it is unnecessary for our purposes), they have a default method and an alternative method with the option \verb|-ampl|. \\

Depending on the options used, the program returns a \verb|.json| file with all the computed information. This file has the following format:

\begin{lstlisting}[language=AMPL]
{
  "generators": [7, 11, 17, 23],
  "frobenius": 27,
  "genus": 17,
  "multiplicity": 7,
  "embedding_dimension": 4,
  "gaps": [1, 2, 3, 4, 5, 6, 8, 9, 10, 12, 13, 15, 16, 19, 20, 26, 27]
}
\end{lstlisting}

\section{Sylvester Denumerant and Numerical Semigroup Membership}

Let $S = \langle a_1,...,a_n \rangle$ be a numerical semigroup. As explained above, the SDP resides in determining if an integer $t \in \mathbb{Z}_{\geq 0}$ is in $S$ and, if the answer is positive, calculate the number of solutions of the linear Diophantine equation
\begin{equation*}
\sum_{i=1}^{n} \lambda_i a_i = t,
\end{equation*} where $\lambda_i \in \mathbb{Z}_{\geq 0}$. As previously stated, we will assume that $i < j$ implies $a_i < a_j$ and that $t \geq a_n$ without loss of generality. We recall that this problem is in NP-hard, and that we do not know a polynomial time algorithm for deciding if a candidate solution is in fact a solution. Therefore, the Sylvester denumerant problem is not known to be in NP, and thus finding a polynomial time algorithm (quantum or not) that solves it would be surprising.\\

In this section we present an algorithm for the calculation of the Sylvester denumerant in the general case. It is a variation of the well-known brute-force approach, with the difference that it gets help from a hypothetical quantum computer. The main idea behind the algorithm is to generate all possible elements of the numerical semigroup up to a certain set of bounds, with the bounds depending on each generator respectively. This generation, which would require an exponential number of calculations in the classical version of the algorithm, can be done in one step inside a quantum computer thanks to the Hadamard gate and quantum paralellism.  \\

First, we shall give an upper bound to all possible solutions for the $\lambda_i$ variables. In order to achieve that, we define 
$$
b_i = 1 + \left\lfloor \log_2 \left( \dfrac{t}{a_i} \right) \right\rfloor,
$$ which tells us the number of binary digits needed for representing all possible values of $\lambda_i$ respectively. Thus, we can calculate 
$$
b = \sum_{i=1}^{n} b_i
$$ 
as the total number of qubits needed. As
$$
\begin{array}{rcl}
b & = & \displaystyle \sum_{i=1}^{n} b_i = \sum_{i=1}^{n} \left(  1 + \left\lfloor \log_2 \left( \dfrac{t}{a_i} \right) \right\rfloor \right) = n + \sum_{i=1}^{n} \left( \left\lfloor \log_2 \left( \dfrac{t}{a_i} \right) \right\rfloor \right) \\ \\
& \leq & \displaystyle n + \sum_{i=1}^{n} \Big( \log_2 (t) - \log_2 (a_i) \Big) = n \big(1 + \log_2(t) \big) - \sum_{i=1}^{n} \log_2 (a_i),
\end{array} 
$$
we can deduce that, in this algorithm, the size of our quantum register grows exponentially with respect to the size of the number of generators of the numerical semigroup, and polynomially with respect of the size of $t$. Thus, fixing the numerical semigroup and just increasing the value of $t$ is not an issue in terms of the computational memory needed, what is really challenging in terms of qubit requirements is adding generators to a numerical semigroup and trying to find the Sylvester denumerant for a fixed $t$. This is not surprising, as the simple task of storing a solution for the aforementioned equation will always take an exponential memory in terms of $n$ (an algorithm that calculates the Sylvester denumerant without finding the actual solutions may exist, though, but it is not this case).

\paragraph{$\mathbb{SETUP}$}
$$$$\noindent\framebox{\parbox[b]{\linewidth}{\begin{algorithmic}
\State $\ket{\psi_0}_{p,b} \leftarrow \ket{0}_p \otimes \ket{0}_b$
\end{algorithmic}}} \\

Our system will have the same setup as in the quantum counting algorithm, where $b$ is the size previously calculated and where $p$ depends on $b$ in such a way that the desired probability of obtaining the correct result is big enough (more on this later).

\paragraph{$\mathbb{STEP}$ 1}
$$$$\noindent\framebox{\parbox[b]{\linewidth}{\begin{algorithmic}
\State $\ket{\psi_1}_{p,b} \leftarrow  \boldsymbol{H}^{\otimes p+n} (\ket{\psi_0}_{p,b})$
\end{algorithmic}}}

The result of this first step is

$$\ket{\psi_1}_{p,b} = \ket{\gamma}_p \otimes \ket{\gamma}_b$$

where

$$
\ket{\gamma}_n = \dfrac{1}{\sqrt{2^n}} \sum_{j=0}^{2^n -1} \ket{j}_n .
$$

\begin{figure}[!b]
\centering
  \begin{tikzpicture}
    \draw (0, 0) rectangle (14, 1);
    \draw (3.5, 0) -- (3.5, 1);
    \draw (7, 0) -- (7, 1);
    \draw (10.5, 0) -- (10.5, 1);
    \node at (1.75,0.5) {$\lambda_1$};
    \node at (5.25,0.5) {$\lambda_2$};
    \node at (8.75,0.5) {$\cdots$};
    \node at (12.25,0.5) {$\lambda_n$};
    \draw (0, 0) -- (0,-0.2);
    \draw (0.875, 0) -- (0.875,-0.2);
    \draw (1.75, 0) -- (1.75,-0.2);
    \draw (2.625, 0) -- (2.625,-0.2);
    \draw (3.5, 0) -- (3.5,-0.2);
    \draw (4.5, 0) -- (4.5,-0.2);
    \draw (7, 0) -- (7,-0.2);
    \draw (5.9, 0) -- (5.9,-0.2);
    \draw (14, 0) -- (14, -0.2);
    \draw (10.5, 0) -- (10.5, -0.2);
    \draw (12, 0) -- (12, -0.2);
    \draw (13, 0) -- (13, -0.2);
    \node at (0.4375, -0.3) {$q_1$};
    \node at (1.3125, -0.3) {$q_2$};
    \node at (2.1875, -0.3) {$\cdots$};
    \node at (12.5, -0.3) {$\cdots$};
    \node at (13.5, -0.3) {$q_{b}$};
    \node at (3.0625, -0.3) {$q_{b_1}$};
    \node at (4, -0.3) {$q_{b_1+1}$};
    \node at (6.5, -0.3) {$q_{b_1+b_2}$};
    \node at (5.25, -0.3) {$\cdots$};
    \node at (11.25, -0.3) {$q_{b_{n-1}+1}$};
  \end{tikzpicture}
\cprotect\caption{Representation of the values of $\lambda_i$ with respect to the $b$ qubits}
\label{fig::qubitregister}
\end{figure}
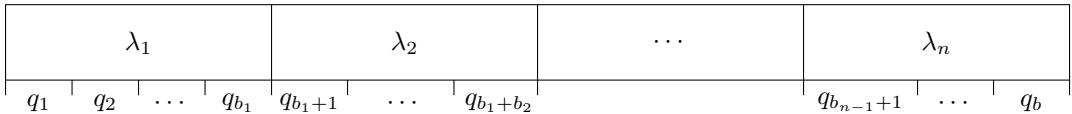

As it is known, thanks to the Hadamard transformation we can obtain all computational basis states of our quantum system inside our register. As the dimension $b$ is calculated with respect to the generators of the numerical semigroup and the integer $t$, these basis states correspond each one uniquely to a combination for all possible values of  the variables $\lambda_i$. A representation of this feature is shown in Figure \ref{fig::qubitregister}, where $q_1,q_2,\ldots,q_{b_1}$ are the $b_1$ binary digits of $\lambda_i$, and so on.

\paragraph{$\mathbb{STEP}$ 2}
$$$$\noindent\framebox{\parbox[b]{\linewidth}{\begin{algorithmic}
\State $\ket{\psi_2}_{p,b} \leftarrow \boldsymbol{C}_{p,b} (\ket{\psi_1}_{p,b})$
\end{algorithmic}}} \\

For the next step, we need the counting gate $\boldsymbol{C}_{p,n}$, defined as 
$$
\boldsymbol{C}_{p,n} : \ket{m}_p \otimes \ket{\psi}_n  \rightarrow \ket{m}_p \otimes (\boldsymbol{G}_n)^m \ket{\psi}_n,
$$ 
where $\boldsymbol{G}_n$ is the Grover gate \cite{grover1996fast}. This Grover gate depends in turn on an Oracle gate $\boldsymbol{O}_{f}$ which yields the following result when applied to a basis state: 
$$
\boldsymbol{O}_{f} : \ket{j}_n \otimes \ket{k}_m \rightarrow \ket{j}_n \otimes \ket{k \oplus f(j)}_m.
$$

We recall that the oracle gate is problem specific, and depends on a certain function $f$ capable of recognising a solution for our problem. What we are going to do is to define $f$ in such a way that it identifies if a certain combination of the variables $\lambda_i$ encoded into our quantum register are a solution for the numerical semigroup membership problem. In other words, if $j$ is the integer codified with $\lambda_1,\ldots,\lambda_n$, then:

\[
  f(j) =
  \begin{cases}
                                   1 & \text{if   $\,\,\,\displaystyle\sum_{i=1}^{n} \lambda_i a_i = t$} \\ \\
                                   0 & \text{otherwise}
  \end{cases}
\]

The rest of the algorithm continues as the counting algorithm in \cite{brassard1998quantum}, but what interests us now is the size of the first register, $p$, which depends on $b$ but also on the probability of obtaining the correct solution for the Sylvester denumerant. Thanks to \cite{boyer1996tight} and \cite{brassard1998quantum}, we know that in order to have a reasonable probability of succeeding, $p \sim \sqrt{2^b}$. Thus, the computational order of the algorithm remains the same in both memory size and time. \\

Concerning the performance of the algorithm, let us compare the iterations required for the brute force approach to calculate the Sylvester denumerant for a fixed semigroup and an increasing value of $t$, against the hypothetical performance of the quantum variant. Let $S = \langle 376,381,393,399 \rangle$ be a numerical semigroup and let $t = 10000$, we can calculate $d(10000; 376,381,393,399)$ with the \verb|numsem| library, which will give us the output seen in Figure \ref{fig::numsemsylvester1}. Thus, we obtain $d(10000; 376,381,393,399) = 9$ and the values of $\lambda_i$ for all nine solutions. \\

\begin{figure}[!tb]
\begin{lstlisting}[language=AMPL]
 d(10000; 376, 381, 393, 399) = 9

 10*(376) + 4*(381) + 12*(393) + 0*(399)
 4*(376) + 13*(381) + 8*(393) + 1*(399)
 10*(376) + 5*(381) + 9*(393) + 2*(399)
 4*(376) + 14*(381) + 5*(393) + 3*(399)
 10*(376) + 6*(381) + 6*(393) + 4*(399)
 4*(376) + 15*(381) + 2*(393) + 5*(399)
 10*(376) + 7*(381) + 3*(393) + 6*(399)
 10*(376) + 8*(381) + 0*(393) + 8*(399)
 16*(376) + 0*(381) + 1*(393) + 9*(399)
\end{lstlisting}
\cprotect\caption{\verb|numsem| output for $d(10000; 376,381,393,399)$}
\label{fig::numsemsylvester1}
\end{figure}

\begin{figure}[!tb]
\includegraphics[width=15cm]{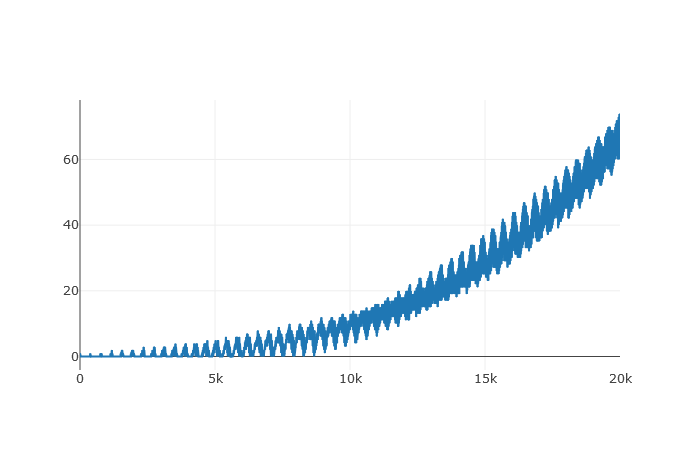}
\centering
\caption{Sylvester quasi-polynomial}
\label{fig::sylv1}
\end{figure}

\begin{figure}[!tb]
\includegraphics[width=15cm]{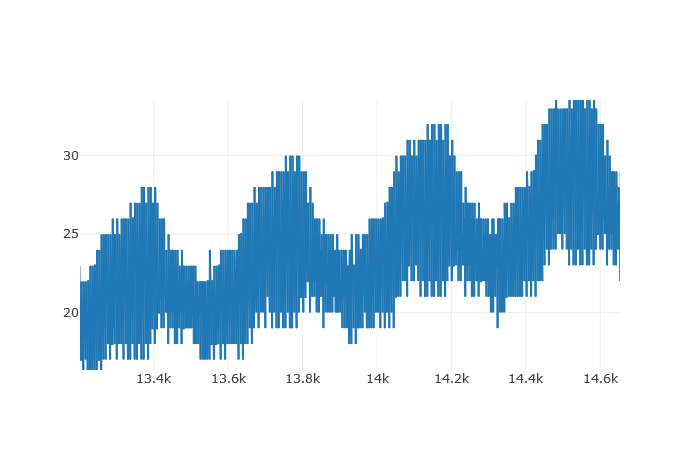}
\centering
\caption{Sylvester quasi-polynomial (detail)}
\label{fig::sylv2}
\end{figure}

\begin{figure}[!tb]
\includegraphics[width=15cm]{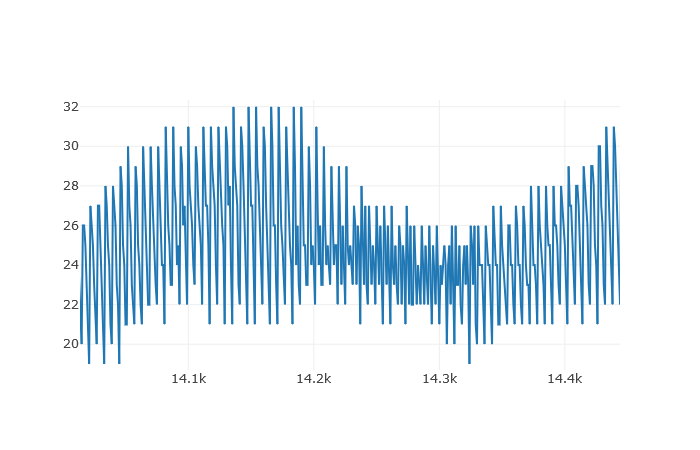}
\centering
\caption{Sylvester quasi-polynomial (detail)}
\label{fig::sylv3}
\end{figure}

What we are going to do know is to calculate $d(t; 376,381,393,399)$ for all values of $t$ up to a certain bound, in order to test the time performance of the brute-force approach and compare it to a hypothetical performance of the quantum approach. First, we can see in Figures \ref{fig::sylv1}, \ref{fig::sylv2} and \ref{fig::sylv3} the Sylvester denumerant function for the numerical semigroup $S = \langle 376,381,393,399 \rangle$ in the interval (1,20000), computed with \verb|numsem| and plotted using Jupyter Notebooks \cite{kluyver2016jupyter}. This function has been proven to be a quasi-polynomial function in the variable $t$ of degree $n-1$ in the general case; however, computing all the coefficients of the polynomial is a computationally hard problem (see \cite{ramirezalfonsin2005diophantine, beck2001polynomial, bell1943interpolated, baldoni2013coefficients}). \\

\begin{figure}[!tb]
\includegraphics[width=15cm]{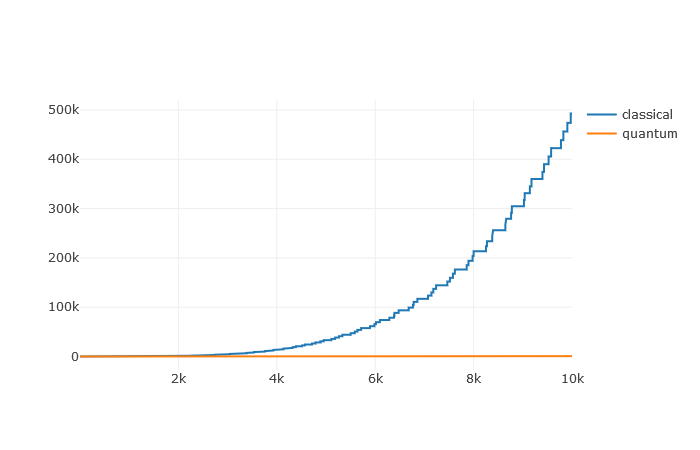}
\centering
\caption{Sylvester denumerant iteration comparison}
\label{fig::sylv-time}
\end{figure}

Additionally, in Figure \ref{fig::sylv-time} we show the operations needed (i.e., the number of possible solutions we have to check) for the computation of $d(t;376,381,393,399)$ with \verb|numsem|, along with the theoretical iterations needed for the quantum version which has been calculated taking into account the asymptotical behavior of the quantum counting algorithm. Please note that both are exponential with respect to $t$, but in the quantum algorithm the exponent would be divided by 2. For example, for $t = 10000$ we would need more than 492,000 classical iterations, and just 702 quantum iterations. However, in the classical case we would obtain the actual solutions for the linear Diophantine equation of the numerical semigroup, while in the quantum case we would \textit{just} obtain the value of the Sylvester denumerant (or one of the possible solutions, if there are any, as will be seen now). \\

Regarding the membership problem of a semigroup $S$, we already know that if $t > f(S)$, then $t \in S$, but the computation of $f(S)$ is extremely hard. Moreover, if $t < f(S)$, the membership problem remains unsolved even if we already know the value of $f(S)$ (although, as proven by Jorge Ramírez-Alfonsín \cite{ramirezalfonsin1996complexity}, it is possible to solve the NSMP in polynomial time if we already have an oracle for the Frobenius problem). \\

The previous quantum algorithm is capable of solving the numerical semigroup membership problem for a certain $t \in \mathbb{Z}_{\leq 0}$. We have to just calculate the Sylvester Denumerant for $t$ and, if $d(S;t) > 0$ return \verb|YES| and otherwise return \verb|NO|. This is quite straightforward but, how about solving the NSMP constructively, not only answering \verb|YES| or \verb|NO| to the question of whether there exist those $\lambda_1, \ldots, \lambda_n$, but also giving one of its possible combinations if the answer is \verb|YES|. For that purpose, we will make use of Grover's Quantum Search algorithm. \\

Using Grover's Algorithm, we can provide a solution for the NSMP linear Diophantine equation in the same $\sqrt{2^b}$ steps as for the SDP. The actual order of both algorithms with respect to $t$, the number of generators $n$ of the numerical semigroup, and the generators $a_1,\ldots,a_n$ is:

$$\sqrt{2^{n \big( 1 + \log_2(t) \big) - \sum_{i=1}^{n} \log_2 (a_i)}}$$

We can also go further and obtain all solutions to the Diophantine equation. First, we run the quantum algorithm for the Sylvester denumerant. Then, we apply the quantum algorithm for the NSMP iteratively until we have obtained as much different solutions as the value of the denumerant. This is a well-known problem in probability theory \cite{feller1957introduction} known as the coupon collector's problem; if the denumerant is $d$, then the expected number of trials $k \geq d$ for obtaining all $d$ solutions grows as $$\Theta(d\log(d)).$$ For example, if $d = 10$, the expected number of runs of the quantum algorithm for the NSMP before obtaining all different solutions is 29.

\section{Conclusions}

In this paper we have shown algorithms for solving the SDP and the NSMP—both related to numerical semigroups—that rely on a hypothetical quantum computer. They are based on the quantum counting and Grover's search algorithms respectively and, as in their counterparts, their complexity can be calculated directly from the maximum number of qubits needed. Thus, as this number has an upper bound of 

$$b\leq n \big(1 + \log_2(t) \big) - \sum_{i=1}^{n} \log_2 (a_i),$$

these algorithms will have a computational order of 
$$\mathcal{O}(\sqrt{2^b}).$$

Note that, although SDP is in principle harder than NSMP, in this approach both solutions share the same complexity.

\bibliographystyle{siam}
\bibliography{refs.bib}

\begin{thebibliography}{10}

\bibitem{baldoni2013coefficients}
{\sc V.~Baldoni, N.~Berline, J.~De~Loera, B.~Dutra, M.~Koeppe, and M.~Vergne},
  {\em Coefficients of {S}ylvester's denumerant}, arXiv preprint
  arXiv:1312.7147,  (2013).

\bibitem{beck2001polynomial}
{\sc M.~Beck, I.~M. Gessel, and T.~Komatsu}, {\em The polynomial part of a
  restricted partition function related to the {F}robenius problem}, The
  Electronic Journal of Combinatorics, 8 (2001), p.~7.

\bibitem{bell1943interpolated}
{\sc E.~Bell}, {\em Interpolated denumerants and {L}ambert series}, American
  Journal of Mathematics, 65 (1943), pp.~382--386.

\bibitem{boyer1996tight}
{\sc M.~Boyer, G.~Brassard, P.~H{\o}yer, and A.~Tapp}, {\em Tight bounds on
  quantum searching}, arXiv preprint quant-ph/9605034,  (1996).

\bibitem{brassard1998quantum}
{\sc G.~Brassard, P.~H{\o}yer, and A.~Tapp}, {\em Quantum counting}, in
  Proceedings of the 25th International Colloquium on Automata, Languages and
  Programming (ICALP), Springer, 1998, pp.~820--831.

\bibitem{feller1957introduction}
{\sc W.~Feller}, {\em An Introduction to Probability Theory and its
  Applications}, 1957.

\bibitem{grover1996fast}
{\sc L.~K. Grover}, {\em A fast quantum mechanical algorithm for database
  search}, in Proceedings of the 28th Annual ACM Symposium on Theory of
  Computing, ACM, 1996, pp.~212--219.

\bibitem{kluyver2016jupyter}
{\sc T.~Kluyver, B.~Ragan-Kelley, F.~P{\'e}rez, B.~E. Granger, M.~Bussonnier,
  J.~Frederic, K.~Kelley, J.~B. Hamrick, J.~Grout, S.~Corlay, et~al.}, {\em
  Jupyter {N}otebooks: a publishing format for reproducible computational
  workflows}, in 20th International Conference on Electronic Publishing, 2016,
  pp.~87--90.

\bibitem{ossoriocastillo2018numsem}
{\sc J.~Ossorio-Castillo}, {\em jqnoc/numsem: numsem console}.
\newblock \url{https://doi.org/10.5281/zenodo.1257967}, 2018.

\bibitem{ossorio2019adiabatic}
{\sc J.~Ossorio-Castillo and J.~M. Tornero}, {\em An adiabatic quantum
  algorithm for the {F}robenius problem}, arXiv preprint arXiv:1907.01789,
  (2019).

\bibitem{papadimitriou1998combinatorial}
{\sc C.~H. Papadimitriou and K.~Steiglitz}, {\em Combinatorial optimization:
  algorithms and complexity}, Courier Corporation, 1998.

\bibitem{ramirezalfonsin1996complexity}
{\sc J.~L. Ram{\'\i}rez-Alfons{\'\i}n}, {\em Complexity of the {F}robenius
  problem}, Combinatorica, 16 (1996), pp.~143--147.

\bibitem{ramirezalfonsin2005diophantine}
\leavevmode\vrule height 2pt depth -1.6pt width 23pt, {\em The Diophantine
  Frobenius problem}, vol.~30, Oxford University Press on Demand, 2005.

\bibitem{rosales2009numerical}
{\sc J.~C. Rosales and P.~A. Garc{\'\i}a-S{\'a}nchez}, {\em Numerical
  semigroups}, vol.~20, Springer Science \& Business Media, 2009.

\bibitem{sylvester1857partition}
{\sc J.~J. Sylvester}, {\em On the partition of numbers}, Quarterly Journal of
  Pure and Applied Mathematics, 1 (1857), pp.~141--152.

\bibitem{sylvester1896outlines}
\leavevmode\vrule height 2pt depth -1.6pt width 23pt, {\em Outlines of seven
  lectures on the partitions of numbers}, Proceedings of the London
  Mathematical Society, 1 (1896), pp.~33--96.

\end{thebibliography}

\end{document}